\documentclass[prx,twocolumn,showpacs,superscriptaddress]{revtex4}
\usepackage{color}
\usepackage{graphicx}
\usepackage{dcolumn}
\usepackage{amsmath}
\usepackage{amssymb}
\usepackage{epstopdf}
\usepackage{subfigure,amsmath,verbatim,moreverb}
\usepackage{multirow}
\usepackage{soul}  
\sethlcolor{yellow}
\usepackage{url}
\usepackage{hyperref}

\definecolor{bluegreen}{rgb}{0,0.2,0.8}
\begin{document}
\title{Energy scaling law for nanostructured materials}
\author{Jianmin Tao}
\altaffiliation{Corresponding author: jianmin.tao@temple.edu \\
URL: \url{http://www.sas.upenn.edu/~jianmint/}}
\affiliation{Department of Physics, Temple University, Philadelphia, 
PA 19122-1801, USA}
\author{Yang Jiao}
\affiliation{Department of Microtechnology and Nanoscience, MC2,
Chalmers University of Technology, Sweden}
\author{Yuxiang Mo}
\affiliation{Department of Physics, Temple University, Philadelphia,
PA 19122-1801, USA}
\author{Zeng-Hui Yang}
\affiliation{Microsystem and Terahertz Research Center, China Academy of
Engineering Physics, Chengdu, Sichuan 610200, China}
\author{Jian-Xin Zhu}
\affiliation{Theoretical Division \& Center for Integrated Nanotechnologies,
Los Alamos National Laboratory, Los Alamos, New Mexico 87545, USA}
\author{Per Hyldgaard}
\affiliation{Department of Microtechnology and Nanoscience, MC2,
Chalmers University of Technology, Sweden}
\author{John P. Perdew}
\affiliation{Department of Physics, Temple University, Philadelphia,
PA 19122-1801, USA}

\date{\today}
\begin{abstract}
The equilibrium binding energy is an important factor in the design of 
materials and devices. However, it presents great computational challenges 
for materials built up from nanostructures. Here we investigate the 
binding-energy scaling law from first-principles calculations. We show 
that the equilibrium binding energy per atom between identical 
nanostructures can scale up or down with nanostructure size. From the 
energy scaling law, we predict finite large-size limits of binding energy 
per atom. We find that there are two competing factors in the determination 
of the binding energy: Nonadditivities of van der Waals coefficients and 
center-to-center distance between nanostructures. To uncode the detail, 
the nonadditivity of the static multipole polarizability is investigated. 
We find that the higher-order multipole polarizability displays 
ultra-strong intrinsic nonadditivity, no matter if the dipole polarizability 
is additive or not. 
\end{abstract}

\maketitle
\section{Introduction}
There is strong interest in nanomaterials, motivated by the development of
nanotechnoloy and by their novel properties arising from quantum confinement.
In particular, the discovery of various atomic-level materials has received
overwhelming attention for their remarkable properties and wide-ranging
applications~\cite{PJena10}. A common feature of these materials is the
strong adhesive van der Waals (vdW) force due to the instantaneous charge
fluctuations. To understand the nature of the vdW force, a variety of
experiments ranging from the smallest atomistic to the largest macroscopic
scales have been performed recently~\cite{Rance10,Loskill13,SRNa14,STsoi14,
Kawai16}. However, details of many surprising phenomena due to the vdW
interaction have not been well understood at the nanoscale~\cite{science16}.
Here we will ask and answer another such question.

The equilibrium binding energy between identical nanostructures is an 
important property involving microscopically the short-range contribution 
arising from the density overlap and the long-range vdW interaction. 
However, due to the large size of nanostructures, it presents great 
computational challenges. As such, an energy scaling law showing the 
variation of equilibrium binding energy per atom with system size is 
highly desired. We here apply an efficient first-principles method, the 
vdW-DF-cx~\cite{KBerland14-1} density functional, to investigate the 
energy scaling law, aiming to provide novel insights into nanostructures. 
Figure 1 shows the energy scaling law for a variety of nanomaterials, 
while Table~\ref{table2} shows the energy scaling law for a variety of 
nanostructures obtained by fitting to our numerical calculations. We find 
that the binding-energy scaling law is largely due to the competing size 
effects of the vdW coefficients and sum of the vdW radii of nanostructures 
determining the intermolecular distance. Our finding is different from 
previous works~\cite{JFDobson06,science16,JFDobson10}, in which the vdW 
coefficients and intermolecular distance are treated independently, 
allowing one to study the dependence of the vdW interaction upon the power 
of distance.

\section{Computational Methods}
The binding energy per atom is defined as $E_b/{\cal N}$, with $\cal N$
being the total number of C/B/N atoms in a nanostructure. For nanotubes,
$\cal N$ is the number of atoms per unit length (in \AA). All our 
calculations of binding energies and distances $d_{cc}$ (center to center) 
and $d_{ww}$ (wall to wall) were performed with Quantum 
ESPRESSO~\cite{QUANTUM-ESPRESSO}, using the vdW-DF-cx~\cite{KBerland14-1} 
nonlocal density functional. We used ultrasoft pesudopotentials with a 
plane-wave energy cutoff at 680 eV. The binding energy $E_b$ is taken as 
the energy difference between relaxed nanostructures in the conventional 
unit cell and in the empty space, except for nanowires, in which the 
distance between two nanowires is fixed at 4.21 \AA~(the relaxed distance 
between C-NWs with seven atoms), due to the high instability. For 
fullerenes, we used $2\times 2\times 2$ for the $k$-mesh. For C-PAHs and 
BN-PAHs, only the $\Gamma$-point  is included in the $k$-mesh, due to the 
large size of the cell. For nanotubes, we used $6\times 6\times 17$ to 
$1\times 1\times 17$ from (3,3) to (40,40). For nanowires, we used 
$3\times 3\times 2$ for all cases.

The nonlocal correlation part of the vdW-DF-cx is the same as in the 
original Rutgers-Chalmers~\cite{Dion04} vdW-DF, which was derived from the
fluctuation-dissipation theorem~\cite{MFuchs02} of electron gas for the
description of the long-range vdW interaction~\cite{PerHyldgaardPRB14}. 
The method contains both the leading-order and higher-order contributions, 
the latter of which are important for solids~\cite{TYR15,ERJ06,ERJ12,
book14,Tao-Rappe14}. Its exchange part is based on a modified semilocal 
functional, which aims to improve the short-range description.

The vdW-DF-cx is a useful first-principles method, as supported by 
Appendices A and B. The experimental geometries of fullerene solids 
(Appendix A) and their equilibrium binding energies (Table I and Appendix B) 
are well reproduced, although $C_6$ for a fullerene pair is 
not, as anticipated in Ref.~\cite{PRL12}. Like SCAN+rVV10~\cite{PengPRX}, 
and with nearly the same binding energy curve, vdW-DF-cx 
predicts~\cite{TT} a chemisorption minimum for graphene on Ni(111) at 
a distance of 2.1~\AA~ from the top nickel plane, in close agreement with 
experiment. Some of us have previously argued~\cite{PerHyldgaardPRB14,
PengPRX} that a vdW functional can be accurate for equilibrium binding 
energies, even if it is not for asymptotic interactions.

To analyse the energy scaling law revealed from our calculation, we 
have to make use of the efficient yet accurate spherical-shell model 
within the single-frequency approximation 
(SFA)~\cite{JTao14,Tao-Rappe16,PRB16} to evaluate the vdW coefficients
between nanostructures. In the SFA, we assume that (i) only valence 
electrons in the outermost subshell are polarizable, and (ii) the density 
is uniform inside the effective radius $R_l$ and zero otherwise. Within 
the SFA, the model dynamic multipole polarizability takes the simple 
expression
\begin{eqnarray}\label{sfapolar}
\label{alpha2}
\alpha_{l}^{\rm SFA}(iu) = R_l^{2l+1}\frac{\omega_l^2}{\omega_l^2+u^2}
~\frac{1-\rho_l}{1-\beta_l\rho_l},
\end{eqnarray}
where $R_l$ is the effective outer radius of the shell,
$\beta_l = \omega_l^2~ {\tilde{\omega}}_l^2/
[(\omega_l^2+u^2)({\tilde{\omega}}_l^2+u^2)]$ describes the coupling of
the sphere and cavity plasmon oscillations, and 
$\rho_l = (1-t_l/R_l)^{2l+1}$ describes the shape of the shell, with $t_l$
being the shell thickness~\cite{GKG04,JTao14}.
$\omega_l = \omega_p \sqrt{l/(2l+1)}$ is the average sphere plasmon 
frequency, ${\tilde{\omega}}_l = \omega_p\sqrt{(l+1)/(2l+1)}$ is the 
cavity plasmon frequency, and $\omega_p = \sqrt{4\pi {\bar n}}$ is the 
average plasmon frequency of the extended electron gas, with 
${\bar n} = N/V_l$ and $V_l$ being the $l$-dependent vdW volume.

Since the geometry effect can be accounted for via $\alpha_l(0)$, the model 
is valid for any geometry~\cite{Tao-Rappe16}. For fullerenes,
$V_l=(4\pi/3) [R_l^3-(R_l-t_l)^3]$. For nanotubes, we take a unit length
to study. The volume per unit length is given as $V_l=2\pi R_0 t_l$, where
$R_0$ is the average radius of a nanotube, which can be accurately
calculated from first principles methods~\cite{PRL06,Duan04,MFerrabone11,
GYGuo07,LWang07}, and $t_l$ is the effective thickness of the nanotube.
For C-NT, we set $t_l=3.4$ bohr, as adopted for fullerene
$V_l=(4\pi/3) [R_l^3-(R_l-t_l)^3]$. 
The volume per unit length is given as $V_l=2\pi R_0 t_l$, where
$R_0$ is the average radius of a nanotube, which can be accurately
calculated from first principles methods~\cite{PRL06,Duan04,MFerrabone11,
GYGuo07,LWang07}, and $t_l$ is the effective thickness of the nanotube.
For C-NT, we set $t_l=3.4$ bohr, as adopted for fullerene
molecules~\cite{GKG04,JTao14}, while for BN-NT, we set $t_l=2.08$
bohr~\cite{MFerrabone11}. For PAH and nanowire, we can make a similar
analysis by taking carbon or BN atoms as a unit. It has been shown that
this model can yield very accurate vdW coefficients~\cite{Tao-Rappe16,
PRB16}.

\begin{figure*}
\includegraphics[width=1.6\columnwidth]{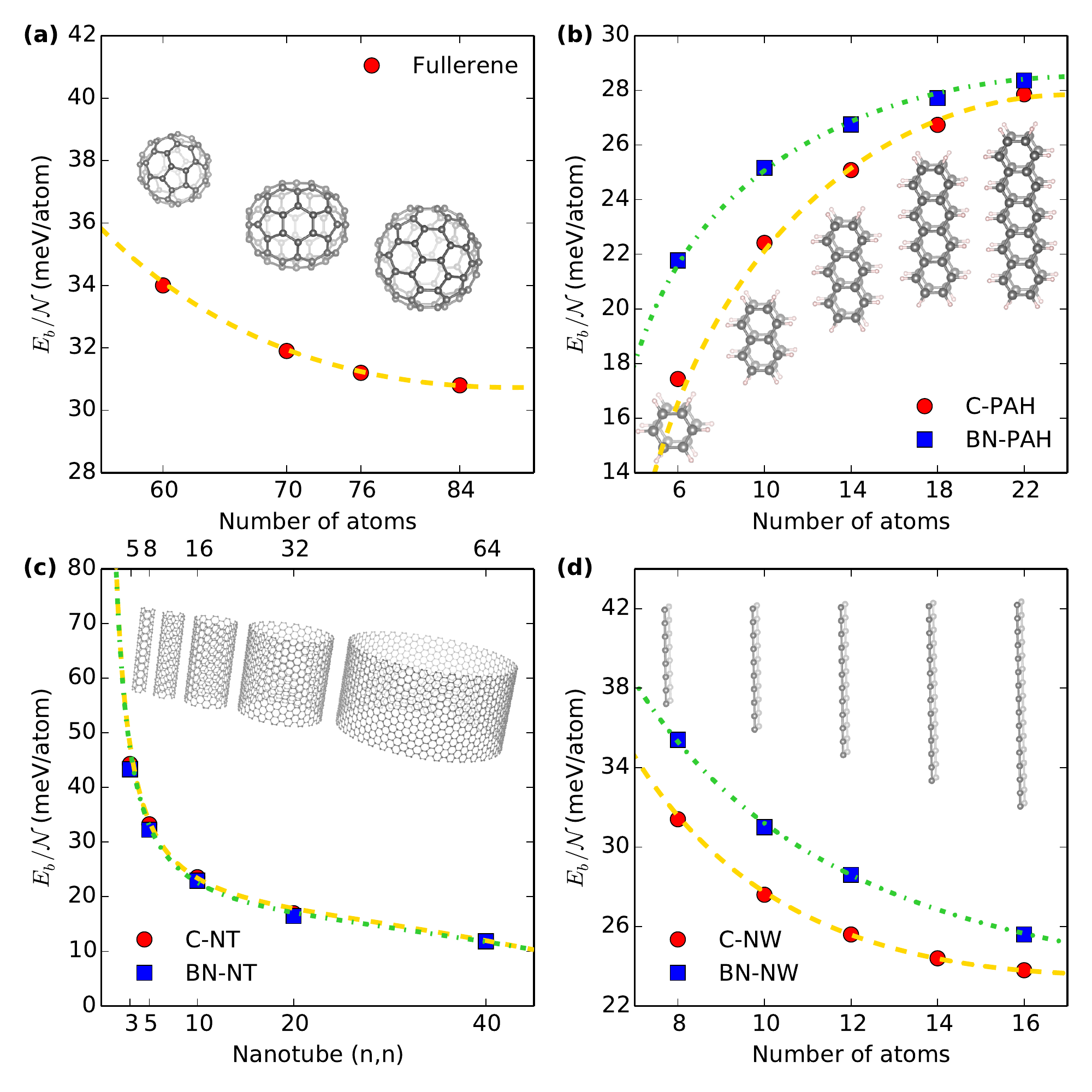}
\caption{Binding-energy scaling law (in meV/atom) of fullerenes in fcc 
solids (Panel (a)), and of pairs of other identical nanoparticles: C-PAHs and 
BN-PAHs in AA stacking dimers (Panel (b)), 
C-NTs and BN-NTs in AA stacking solids (Panel (c)), as well as carbon 
nanowires (C-NWs) and boron-nitride substitutes (BN-NWs) in AA stacking 
dimers (Panel (d)), with system size. $\cal N$ on the vertical axis is the 
total number of C/B/N atoms in a nanostructure. The dotted curve 
(yellow for carbon and green for BN) is the analytic modeling from 
Table~\ref{table2}.}
\label{figure1}
\end{figure*}

\section{Binding-energy scaling law for nanostructures}
{\bf Ball-to-ball interaction.}
Fullerene is an important class of nanomaterials with a variety of
applications~\cite{RLoutfy02} in physics, chemistry, and materials
science. The binding energy between fullerene pairs is a typical example
of ball-ball interaction~\cite{per11,TYR15}. We have calculated the
binding energies per atom of fullerene solids with the optimized 
fcc-type geometries (Appendix A).  
For C$_{60}$, the experimental value was already extrapolated
to 0 K, while for others, the experimental values are available only at
high temperature ($\sim 850$ K)~\cite{MMoalem95}, due to the strong vdW
force. We have estimated the thermal energy correction
($\sim 2$ meV/atom) for C$_{70}$ solid, for which the heat capacity data
are available~\cite{MMoalem95}. For C$_{76}$~\cite{BBrunetti97} and 
C$_{84}$~\cite{MMoalem95}, we take the same 
thermal correction per atom as for C$_{70}$. From Table~\ref{table1}, we observe
that the calculated binding energies are generally within the range of
thermally corrected experiments, while the intermolecular distance $d_{cc}$
is between the DFT-LSDA~\cite{TYR15} (local spin-density approximation)
and experimental value~\cite{Magomedov05} (at room temperature). This
suggests that the method used in this work is not only reliable for binding
energy, but also for the center-to-center distance. See appendix A for
further detailed discussion.

Figure 1(a) shows that the binding energy of fullerenes per atom decreases
slowly (only 3 meV/atom from C$_{60}$ to C$_{84}$ ) with fullerene size,
while Table~\ref{table2} shows that the large-size limit of binding energy
saturates to a constant 24 meV/atom. From Table~\ref{table1}, we can see 
that the center-to-center distance $d_{cc}$ gradually becomes larger and 
larger from C$_{60}$ to C$_{84}$. With the increase of fullerene size, vdW
coefficients per atom pair, in particular the higher-order ones
(Table~\ref{table3}), dramatically increase, while the sum of the vdW radii
of fullerenes characterized by $d_{cc}$ also increases. These two factors have 
opposite effects on the energy scaling law, leading to the slow variation of 
the binding energy with system size. It is interesting to note that the 
wall-to-wall distance $d_{ww}$ of fullerenes gets slightly shrunk from C$_{60}$ 
to C$_{84}$. This is because the vdW force between fullerenes increases 
from C$_{60}$ to C$_{84}$ , pulling two fullerenes slightly closer.

\begin{table*}
\caption{Variation of binding energies per atom $E_b/{\cal N}$
(in meV/atom) of fullerenes in fcc solids, and of pairs of other identical 
nanoparticles: C-PAHs and BN-PAHs, C-NTs and BN-NTs, and
carbon and BN nanowires with system size, and center-to-center ($d_{cc}$) and
wall-to-wall ($d_{ww}$) distances (\AA). We extrapolated experimental
binding energies~\cite{MMoalem95} at 850 K to 0 K for fullerenes, except
for C$_{60}$, which was already extrapolated to 0 K, by estimating the
thermal correction per atom from the thermal heat capacity of 
C$_{70}$~\cite{MMoalem95} and then making the same correction per atom 
from C$_{70}$ to the experimental values for fullerenes 
C$_{76}$~\cite{BBrunetti97} and C$_{84}$~\cite{MMoalem95}. 
Experimental center-to-center distances are from 
Ref.~\cite{Magomedov05}. All $d_{cc}$ between wires are fixed at 4.21 \AA.
}
\begin{tabular}{lccccc|cc}
\hline \hline
  &   &${\cal N}$ & $d_{cc}$ (\AA) & $d_{ww}$ (\AA) &
$E_b^{\rm cal}/{\cal N}$ &$d_{cc}^{\rm expt}$ (\AA) & $E_b^{\rm expt}/{\cal N}$\\
\hline
\multirow{5}{2.0cm}{Fullerene}
& C$_{60}$(Ih)             &60&  9.96  & 3.05  & 35.0 &10.02      &28 $\sim$ 35 \\
& C$_{70}$(D$_{5h}$)       &70& 10.52  & 2.99  & 32.9 &10.61      &26 $\sim$ 32 \\
& C$_{76}$(D$_{2}$)        &76& 10.92  & 2.92  & 32.2 &10.94      &28 $\sim$ 29 \\
& C$_{84}$(D$_{2}$)        &84& 11.06  & 2.78  & 31.6 &11.36      &29 $\sim$ 35 \\
\hline
\multirow{5}{2.0cm}{C-PAH}
& Benzene                  &6 & 4.07    &--     & 17.4 &\\
& Naphthalene              &10& 3.95    &--     & 22.4 & \\
& Anthracene               &14& 3.90    &--     & 25.0 &\\
& 2,3-Benzanthracene       &18& 3.86    &--     & 26.8 &\\
& Pentacene                &22& 3.84    &--     & 27.8 &\\
\hline
\multirow{5}{2.0cm}{BN-PAH}
& B$_3$N$_3$H$_6$          &6 & 3.98    &--     & 21.8 &\\
& B$_5$N$_5$H$_{8}$        &10& 3.90    &--     & 25.2 &\\
& B$_7$N$_7$H$_{10}$       &14& 3.85    &--     & 26.8 &\\
& B$_9$N$_9$H$_{12}$       &18& 3.84    &--     & 27.6 &\\
& B$_{11}$N$_{11}$H$_{14}$ &22& 3.83    &--     & 28.4& \\
\hline
\multirow{5}{2.0cm}{C-NT}
&  (3,3)                   &5 &7.28   & 3.08  & 44.3 &\\
&  (5,5)                   &8 &10.01  & 3.14  & 33.2 &\\
&(10,10)                   &16&16.78  & 3.15  & 23.6 &\\
&(20,20)                   &32&30.34  & 3.15  & 17.0 &\\
&(40,40)                   &64&57.47  & 3.15  & 11.9 &\\
\hline
\multirow{4}{2.0cm}{BN-NT}
& (3,3)                    &5 &7.18   & 2.95  & 43.3 &\\
& (5,5)                    &8 &10.07  & 3.10  & 32.3 &\\
& (10,10)                  &16&16.97  & 3.11  & 22.9 & \\
& (20,20)                  &32&30.78  & 3.10  & 16.5 &\\
& (40,40)                  &64&58.38  & 3.10  & 11.8 &\\
\hline
\multirow{5}{2.0cm}{Carbyne (C$_N$)}
& C$_8$                    &8 & 4.21   &--     &31.4  & \\
& C$_{10}$                 &10& 4.21   &--     &27.7  & \\
& C$_{12}$                 &12& 4.21   &--     &25.6  & \\
& C$_{14}$                 &14& 4.21   &--     &24.4  & \\
& C$_{16}$                 &16& 4.21   &--     &23.8  & \\
\hline
\multirow{5}{2.0cm}{BN-carbynes (BN)$_{N/2}$}
& (BN)$_4$                 &8 & 4.21   &--     &35.4  & \\
& (BN)$_{5}$              &10& 4.21   &--      &31.0  & \\
& (BN)$_{6}$              &12& 4.21   &--      &28.6  & \\
& (BN)$_{8}$              &16& 4.21   &--      &25.6  & \\
\hline \hline
\end{tabular}
\label{table1}
\end{table*}

{\bf Plane-to-plane interaction.}
Polycyclic aromatic hydrocarbons (C-PAHs) are a large class of conjugated
$\pi$-electron systems of great importance in many areas such as
environmental chemistry, materials science, and astrochemistry~\cite{Ref1}.
The energy scaling law between C-PAHs reflects the plane-to-plane vdW
interaction~\cite{PHyldgaard10,Rangel16,Podeszwa10}. Here we focus on the 
binding energies per atom of C-PAH dimers with the optimized AA stacking.
Crucial to this problem is the fact that the center-to-center distance
$d_{cc}$ remains nearly the same for all C-PAH, with a slight decreasing
trend similar to $d_{ww}$ for fullerene pairs, as shown in Table~\ref{table1}.
This is because from benzene to pentacene, the vdW force increases, pulling
two planar molecules slightly closer. Since the vdW coefficients per atom
rapidly increase with system size~\cite{MALMarques07}, due to the nonadditivity
arising from $\pi$-electron delocalization, while their center-to-center
distance $d_{cc}$ does not change much, the binding energy between C-PAHs
scales up rapidly, as shown in Fig. 1(b). A similar energy scaling law is
also observed for boron-nitride (BN) substitues~\cite{Ref2} for the same
reason, as shown in Table~\ref{table1} and Fig. 1(b), respectively. 
From Table~\ref{table2}, we see that our energy scaling law
predicts the same binding energy 30 meV/atom between two identical long-chain 
limits of PAH and BN-PAH (AA stacking). Note that this limit is physically
different from a bi-layer of infinite two-dimentional sheets.

{\bf Tube-to-tube interaction.}
Carbon nanotubes (C-NTs) are perhaps one of the most-widely studied
nanomaterials, due to their many unusual properties and 
applications~\cite{Ref3}. Study of
their energy scaling law is of broad interest. A C-NT has cylindrical
symmetry. It is characterized by a pair of integer parameters $(n,m)$,
with radius given by $(\sqrt{3}a/2\pi)\sqrt{n^2+m^2+nm}$, with $a$ being
the bond length. When $n=m$, it takes the armchair structure, while for
$n \neq m$, it takes the zigzag structure. Their size can be adjusted with
$n$ or $m$. Here we focus on the binding energies per atom of the optimized
close-packed solids of infinitely-long armchair C-NTs. Figure 1 (c) shows
the variation of binding energy per atom with tube size for the AA stacking 
at the optimized geometry. From Fig. 1(c), we observe that
when the size of C-NT increases from $(3,3)$ to $(40,40)$, the binding
energy per atom drops significantly from 44.3 meV/atom to 11.9 meV/atom.
Table~\ref{table1} shows that the wall-to-wall distance $d_{ww}$ is nearly
a constant with tube size, while the center-to-center distance $d_{cc}$
dramatically increases, a situation similar to fullerene. This largely
decreases the vdW force, due to the fact that the nonadditivity of vdW
coefficients is unable to cancel that of the vdW radii for C-NT pairs,
leading to the decreasing trend of the binding energy with tube size.
Clearly, this trend has been followed by BN-NTs as shown in Fig. 1(c) and
Table~\ref{table1}. However, the binding energy for BN-NTs is slightly
smaller than that for C-NTs. A possible explanation is that, because C-NTs
and BN-NTs take structures similar to those of their bulks (the BN layered
materials), the BN atoms in BN-NTs may not be all on the same surface, as
they are for C-NTs. This will increase the band gap (5.5 eV)~\cite{XBlase94}
of BN-NT (a situation similar to h-BN~\cite{GCassabois16}) and thus decrease
the vdW coefficients between BN-NTs, compared to C-NTs, as shown in
Table~\ref{table3}. The energy scaling law in Table~\ref{table2} predicts
the binding energies of 12.8 meV for C-NT, and 12.0 meV for BN-NT with AA
stacking in the large size limit, which are rather close to 11.5 meV/atom
for graphene (quantum Monte Carlo value)~\cite{EM15} and 9.9 meV/atom of
h-BN~\cite{GCPRL13}.

{\bf Wire-to-wire interaction.}
Carbyne is a carbon-based nanowire (C-NW) with an infinite chain of
$sp$-hybridized carbon atoms, held together by either double or
alternating single and triple atomic bonds. It displays unusual
properties, such as strong chemical activity and
extreme instability in ambient conditions. C-NW and its BN substitute (BN-NW)
have attracted great attention recently~\cite{LShi16,MLiu13,OCretu14,
science16}, due to a variety of remarkable properties. Here we study
the variation of binding energy per atom between two AA-stacked
finite-length C-NWs with system size. Due to the instability of C-NW, the
binding energy is calculated at a fixed distance between two C-NWs, rather
than at the relaxed distance (see Computational Methods for detail). 
As shown by Fig. 1(d), the binding energy per atom between C-NWs decreases 
with system size. This is rather similar to those of fullerenes and 
nanotubes, but with much stronger size-dependence. It is also opposite to 
the energy scaling law of PAHs. This feature has been inherited by its 
BN substitute. However, the binding energy between BN-NWs is slightly 
stronger, due to the additional permanent dipole-dipole interaction 
between B and N atoms, a similar situation to BN-PAHs. In the large-size 
limit, the binding energy (16 meV/atom) between C-NWs becomes slightly 
larger than that (15.2 meV/atom) between BN-NWs, suggesting that 
crossover arising from the distortion of BN atoms occurs somewhere.

\begin{table}
\caption{
Binding-energy scaling law (in meV), $E_b/{\cal N} = a + b/{\cal N} + 
1000c[({\cal N}-d)^2/{\cal N}]/\{1 + [c({\cal N}-d)]^4\}$, for nanostructured
materials, where the parameters characterized by specific nanostructures
are determined by a fit to numerical binding energies in
Table~\ref{table1}. }
\begin{tabular}{l|ccccc}
\hline \hline
               &$a$   & $b$     & $c$        & $d$ \\
Fullerene pair &24    &550      &0.00025     &75   \\
C-PAH pair     &30    &$-50$    &$-0.00002$  &22   \\
BN-PAH pair    &30    &$-35$    &$-0.00001$  &22   \\
C-NT pair      &12.8  &170      &$-0.0000018$&20 \\
BN-NT pair     &12    &170      &$-0.0000015$&20 \\
C-NW pair      &16    &115      &0.0006      &12 \\
BN-NW pair     &15.2  &160      &0.0002      &10 \\
\hline \hline
\end{tabular}
\label{table2}
\end{table}

\section{Discussion}
To understand the energy scaling law of nanostructures, knowledge of
the vdW coefficients is essential. Due to the direct relevance of the vdW
coefficients to the static multipole polarizability, and in view of the
relatively large size of the nanostructures, our starting point is the 
classical conducting solid or hollow sphere model, which is exact for 
slowly-varying densities. In this model, the static multipole 
polarizability satisfies~\cite{PRL12}
\begin{eqnarray}\label{eq1} 
\label{alpha}
\alpha_l(0) = [\alpha_1(0)]^{(2l+1)/3},
\end{eqnarray}
where $l$ is the order of the polarizability, with $l=1$ (dipole), 2
(quadrupole), 3 (octupole), etc. (The dipole polarizability of a fullerene
can be estimated~\cite{GKG04} from $\alpha_1(0) = [R_N+t/2]^3$, where
$R_N$ is the average radius of the nuclear framework of a fullerene, and
$t$ is the effective thickness of the shell.) Suppose the sphere contains
${\cal N}_i$ identical atoms with the static multipole polarizability
$\alpha_l^i(0)$. Now let the volume of the sphere increase from $V_i$ to
$V_f$ with fixed electron density, so that ${\cal N}_i$ will increase to
${\cal N}_f$. We seek interpolation relating the multipole
polarizabilities at ${\cal N}_i$ and ${\cal N}_f$, the endpoints of the
range over which we know $\alpha_l(0)$. The dipole polarizability per atom
at ${\cal N}_i$ can be written as
$\alpha_1^i(0)/{\cal N}_i = \alpha_1^f(0)/{\cal N}_f^{(1+\delta_1)}$.
If $\delta_1=0$, the dipole polarizability is additive. Otherwise, it is
nonadditive~\cite{MScheffler12}. Similarly, the higher-order
polarizabilities can be written as
\begin{eqnarray}\label{eq3}
\alpha_l^i(0)/{\cal N}_i &=& \alpha_l^f(0)/{\cal N}_f^{(1+\delta_l)},
\end{eqnarray}
where $\delta_l$ is a measure of nonadditivity of the multipole
polarizability. Substituting Eq.~(\ref{eq1}) into Eq.~(\ref{eq3}) and
performing some simple algebra, we can express the nonadditivity of the
multipole polarizability in terms of that of the dipole polarizability as
\begin{eqnarray}\label{eq4}
\delta_l= 
[(2l+1)(1+\delta_1)-3]/3-\frac{1}{3}[(2l+1)-3]\bigg(\frac{{\rm ln} 
{\cal N}_i}{{\rm ln} {\cal N}_f}\bigg).
\end{eqnarray}
The last term of Eq.~(\ref{eq4}) is the size correction to the nonadditivity
of the higher-order multipole polarizability. 
It vanishes for $l=1$ (dipole), and in the classical limit
(${\cal N}_f \rightarrow \infty$).
When the dipole polarizability is additive (i.e., $\delta_1=0$), we can
still observe the strong nonadditivity of the higher-order multipole
polarizability (i.e., $\delta_l >0$). Therefore, the nonadditivity
of the higher-order multipole polarizability is an intrinsic property of a
material. For any ${\cal N}_i < {\cal N} < {\cal N}_f$, we just replace
${\cal N}_f$ by ${\cal N}$ in Eq.~(\ref{eq4}). For the smallest possible
data set ${\cal N}_i={\cal N}_f$, our formulas would predict $\delta_l=0$.
In other words, the multipole polarizability at a single point $\cal N$
cannot deliver the physical nonadditivity. In order to identify the
physical nonadditivity in which the initial value should be taken from an
atom~\cite{JTao14}, we need the multipole polarizability at $\cal N$ or
${\cal N}_f$ that is reasonably larger than ${\cal N}_i$. The scaling 
properties for the static multipole polarizabilities of various 
nanostructures are given by Table~\ref{table3}. [The dipole 
polarizabilities~\cite{Mark92} of nanotubes are taken from 
Refs.~\cite{PRL06,Duan04,MFerrabone11,GYGuo07,LWang07}, while the 
higher-order polarizabilities are estimated from Eq.~(\ref{eq1}).]

\begin{table*}
\caption{
Scaling properties of the static multipole polarizabilities of
fullerenes, C-NTs, and BN-NTs, with armchair (m,m) and zigzag
(m,0) structures, and  the vdW coefficients between identical pairs.
${\cal N}$ is the number of atoms in a nanostructure. For
NTs, it represents the number of atoms per unit length ($\AA$). The static
dipole polarizability and $C_6$ for fullerenes are the {\em ab initio}
values taken from Ref.~\cite{JCP13}, while the higher-order static
polarizabilities and vdW coefficients are calculated in this work. For
nanotubes, the static dipole polarizabilities are the {\em ab initio} values
taken from the literature~\cite{PRL06,Duan04,MFerrabone11,GYGuo07,LWang07},
while all others are obtained in this work. }
\begin{tabular}{l|l|lcc}
\hline \hline
                      && Polarizability scaling  \\ 
\hline
Fullerenes (present)  &&$\alpha_1(0)/{\cal N}^{1.19}$&$\alpha_2(0)/{\cal N}^{1.65}$ &
$\alpha_3(0)/{\cal N}^{2.11}$  \\
Fullerenes ({\em ab initio}) & &$\alpha_1(0)/{\cal N}^{1.2}$& & \\
\hline
\multirow{2}{60pt}{Carbon nanotubes}  &Armchair (m,m)
&$\alpha_1(0)/{\cal N}^{1.27}$  &$\alpha_2(0)/{\cal N}^{1.77}$ &
$\alpha_3(0)/{\cal N}^{2.26}$ \\ \cline{2-2}
&zigzag (m,0) &
$\alpha_1(0)/{\cal N}^{1.16}$  &$\alpha_2(0)/{\cal N}^{1.62}$ &
$\alpha_3(0)/{\cal N}^{2.09}$ \\ \cline{2-2}
\hline
\multirow{2}{60pt}{BN-based nanotubes} &Armchair (m,m)
&$\alpha_1(0)/{\cal N}^{1.06}$  &$\alpha_2(0)/{\cal N}^{1.59}$ &
$\alpha_3(0)/{\cal N}^{2.13}$ \\ \cline{2-2}
&zigzag (m,0) &
$\alpha_1(0)/{\cal N}^{1.06}$  &$\alpha_2(0)/{\cal N}^{1.42}$ &
$\alpha_3(0)/{\cal N}^{1.76}$ \\
\hline
   && vdW coefficients' scaling \\
\hline\hline
Fullerenes (present work) &  &$C_6/{\cal N}^{2.26}$  &$C_8/{\cal N}^{2.73}$ &
$C_{10}/{\cal N}^{3.2}$  \\
Fullerenes ({\em ab initio}) & &$C_6/{\cal N}^{2.20}$  & &  \\
\hline
\multirow{2}{60pt}{C-NTs}  &Armchair $(n,n)$
&$C_6/{\cal N}^{3.10}$  &$C_8/{\cal N}^{4.08}$ &
$C_{10}/{\cal N}^{5.08}$ \\ \cline{2-2}
&zigzag $(n,0)$ &
$C_{6}/{\cal N}^{2.66}$  &$C_{8}/{\cal N}^{3.60}$ &
$C_{10}/{\cal N}^{4.54}$ \\ \cline{2-2}
\hline
\multirow{2}{60pt}{BN-NTs} &Armchair $(n,n)$
&$C_{6}/{\cal N}^{2.26}$  &$C_{8}/{\cal N}^{2.98}$ &
$C_{10}/{\cal N}^{3.68}$ \\ \cline{2-2}
&zigzag $(n,0)$ &
$C_{6}/{\cal N}^{2.12}$  &$C_{8}/{\cal N}^{3.14}$ &
$C_{10}/{\cal N}^{4.16}$ \\
\hline\hline
\end{tabular}
\label{table3}
\end{table*}

With the scaling properties of the static multipole polarizability, we can 
study the nonadditivity of the vdW interactions. The vdW coefficients 
between two identical solid spheres, each having ${\cal N}$ atoms, take 
the simple form~\cite{JTao14}
$C_6 = \alpha_1(0)\alpha_1(0) h_6({\bar n})$,
$C_8 = \alpha_1(0)\alpha_2(0) h_8({\bar n})$, and
$C_{10} = \alpha_1(0)\alpha_3(0) h_{10,1}({\bar n})+
\alpha_2(0)\alpha_2(0) h_{10,2}({\bar n})$.
Here ${\bar n}$ is the average valence electron density of the sphere,
and $h_{6}$-$h_{10}$ are functions of ${\bar n}$ determined by the
Casimir-Polder formula. Our calculation shows that ${\bar n}$ is nearly
a constant with system size, so that the nonadditivity of vdW coefficients
is essentially determined by the nonadditivity of the static multipole
polarizability. According to Eqs.~(\ref{eq1})-(\ref{eq4}), we can express
the nonadditivity of the vdW coefficients as
\begin{eqnarray}\label{eq5}
C_6^i/{\cal N}_i^2 &=& C_6^f/{\cal N}_f^{2+2\delta_1}.
\end{eqnarray}
If $\delta_1=0$ or the dipole polarizability is additive, so is $C_6$.
Similarly, we have
\begin{eqnarray}
\label{eq6}
C_8^i/{\cal N}_i^2 = C_8^f/{\cal N}_f^{2+\delta_1+\delta_2},~~ 
C_{10}^i/{\cal N}_i^2 = C_{10}^f/{\cal N}_f^{2+\delta_1+\delta_3}, 
\end{eqnarray}
where $\delta_l$ are given by Eq.~(\ref{eq4}). (Note that
$\delta_1+\delta_3=2\delta_2$.) From Eqs.~(\ref{eq4})-(\ref{eq6}), we can
see that the nonadditivity of the vdW coefficients
($2\delta_1$ for $C_6$,
$\delta_1+\delta_2$ for $C_8$, and $\delta_1+\delta_3$ for $C_{10}$)
largely arises from that of the multipole polarizability. If $C_6$ is
additive, $C_8$ and $C_{10}$ are still nonadditive, because, even if
$\delta_1=0$, $\delta_2$ and $\delta_3$ are not zero. This finding suggests
that the nonadditivity of higher-order vdW coefficients essentially
originates from the intrinsic nonadditivity of the multipole polarizability.
The scaling properties for the vdW coefficients between nanostructures are
also listed in Table~\ref{table3}.

With knowledge of the nonadditivity of vdW coefficients, we can now explain
the energy scaling law for fullerenes as follows. Let us consider the
interaction between two identical classical solid spheres that are close
enough~\cite{Perdew12}. Each sphere has a radius of
$R = [\alpha_1(0)]^{1/3}$. The vdW coefficients are
\begin{eqnarray}\label{eq8}
C_6 \sim \alpha_1(0)\alpha_1(0) \sim R^6 \sim {\cal N}^{2(1+\delta_1)},
 \\
C_8 \sim \alpha_1(0)\alpha_2(0) \sim R^8 \sim {\cal N}^{8(1+\delta_1)/3},
 \\
C_{10} \sim \alpha_1(0)\alpha_3(0) \sim R^{10} 
\sim {\cal N}^{10(1+\delta_1)/3}.
\end{eqnarray}
This yields
\begin{eqnarray}\label{eq8p}
C_{2j}/(2R)^{2j} \sim R^0.
\end{eqnarray}
When the spheres are close to each other, all energy terms of the vdW
series are independent of $R$ or system size. This nonadditivity
cancellation is valid for both solid spheres and hollow spheres with a
cavity, because our analysis for solid spheres is also valid for
hollow spheres. To demonstrate this cancellation, we have calculated
the energy series $C_{6}/d^{6}$, $C_{8}/d^{8}$ and $C_{10}/d^{10}$ for
fullerene solids, with $d=d_{cc}$ given in Table~\ref{table1}. Our
calculation shows that both the leading-order and higher-order energy
terms are nearly size-independent. However, it was found~\cite{Perdew12} 
that this series diverges when two identical classical conducting spheres 
touch, but this spurious divergence can be removed without changing the 
asymptotic series. Nevertheless, the binding energy per atom 
($[C_{2j}/(2R)^{2j}]/{\cal N}$) is decreasing with ${\cal N}$, as shown 
by Fig. 1(a).

The physics behind the energy scaling law is now clear. The behavior
observed in Fig 1(a)-1(d) is a consequence of a competition between the
nonadditivities in the vdW coefficients and in the vdW radii, which are
saturated to the bulk values. For ball-ball interactions, there is
large cancellation between $C_{2j}$ and $(2R)^{2j}$, leading to
a rather slow variation of the binding energy per atom with system size.
For plane-plane interactions, the nonadditivity of vdW coefficients is
dominant, because the parallel distance $d_{cc}$ is nearly a constant,
leading to significant increase in binding energy. For tube-tube
interactions, the nonadditivity of vdW coefficients becomes relatively
less important due to the much larger size of tubes, compared to that
of fullerenes, leading to a faster variation of the binding energy per
atom with system size than that between fullerenes. There is a difference
between BN atoms in BN-PAH and BN-NT. The reason is that BN atoms of
BN-PAH can form $\pi$-electron delocalization, as in C-PAH.
However, as in bulk h-BN, BN atoms of BN-NT can not, because both BN-NT
and bulk h-BN have large gaps~\cite{GCassabois16,XBlase94}. This
difference in bonding explains why the binding energy of BN-PAHs is
greater than that of C-PAHs, but the binding energy for BN-NTs is slightly
smaller than that for C-NTs. From BN-NW to h-BN bulk material, we
can see that the energy gap evolves from a small value (nanowire) to
a larger value 5.5 eV (BN-NT), to a even larger value 5.9 ev (h-BN),
suggesting the deformation of B atoms from the surface of N atoms and thus
a change in energy scaling law from nanowires to nanotubes. Due to the
$\pi$-electron delocalization, the difference in binding energy scaling
between C-PAH and BN-PAH is the same as that between C-NW and
BN-NW: BN-based PAHs and BN-based nanowires show faster energy
variation with system size than C-based counterparts. The binding energy
per atom of the nanowire dimers may decrease with increasing length due to
dilution of the effect of covalent bonding between C atoms at the ends of
the dimer.

\section{Conclusion}
The binding energy determines the stability of nanostructures and is 
therefore very important in the study of nanostructures. However, it has 
presented computational challenges. In this work, we have studied the 
binding energy law of nanostructures based on a first-principles method. 
We find that there is a binding-energy scaling law between identical 
nanostructures. From the law, we can predict the binding energy at any 
structure size. We illustrate this finding with fullerenes, PAHs, 
nanotubes, and nanowires. Apart from fullerenes, we chose AA stacking in 
our study. From the energy scaling law, we predict finite large-size 
limits, which are expected. To understand the energy-scaling law, we have 
studied the vdW coefficients using the accurate hollow-sphere model within 
the SFA. We find that the energy scaling law is determined by two 
competing factors: Nonadditivities of the vdW coefficients and the 
center-to-center distance. This leads us to conclude that the 
energy-scaling law in part originates from the nonadditivity of the 
static multipole polarizability of nanostructures.

\section{Acknowledgements}
The authors thank Mark R. Pederson and Roberto Car for valuable comments
and suggestions, Jing Yang, Guocai Tian, and Haowei Peng for useful
discussions and technical help, and Hong Tang for useful comments. YM 
acknowledges support from the NSF under Grant No. CHE 1640584. JT was 
supported by the DOE under grant No. DE-SC0018194. JT was also supported 
on Temple start-up from John P. Perdew. YJ and PH acknowledge support by 
the Swedish Research Council (VR) and the Swedish Research Foundation (SSF)
under contract SE13-0016.  ZY and JPP were supported by the NSF 
under Grant No. DMR-1607868. ZY was also supported by Science Challenge 
Project No. TZ2016003 (China). JXZ acknowledges the support by the Center 
for Integrated Nanotechnologies, a DOE BES user facility. Computational 
support was provided by the HPC at Temple University and by the Swedish 
National Infrastructure for Computing through allocations at HPC2N 
(Ume{\aa}) and C3SE (Gothenburg).

\appendix
\section*{APPENDICES}

\section{vdW-DF-cx characterization of Fullerenes}
 
\setcounter{figure}{0}
\renewcommand{\thefigure}{A\arabic{figure}}
\begin{figure}
\centering
\includegraphics[width=0.9\columnwidth]{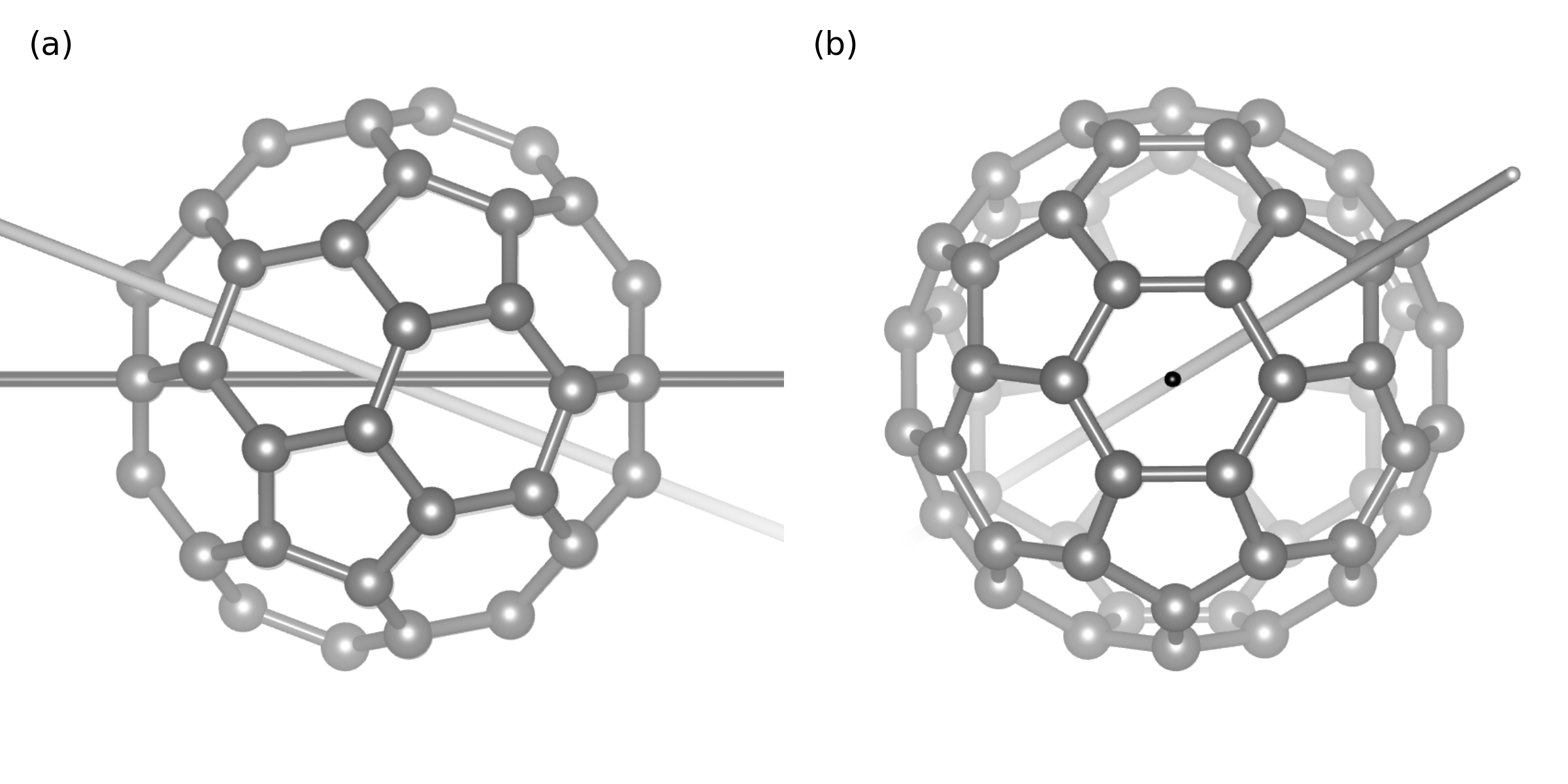}
\caption{The atomic configuration of C$_{60}$ as seen perpendicular to
(left panel) and along (right panel) the major symmetry axis, shown as
a dark gray line in the left panel. The atomic configurations of
C$_{70}$, C$_{76}$ and C$_{80}$ fullerenes are shown as inserts in Fig.\ 1a 
of the main text. The major symmetry axis of C$_{60}$ goes through a
pair of hexagonal facets and it is experimentally found to be aligned 
with the $[111]$ direction of the C$_{60}$ fcc crystal, below 
260 K \cite{David92}. The C$_{60}$ molecule also has a secondary 
symmetry line (lighter gray line in both panels) which goes 
though a pair of pentagonal facets.
\label{fig:C60axis}}
\end{figure}

\renewcommand{\thefigure}{A\arabic{figure}}
\begin{figure}
\centering
\includegraphics[width=0.8\columnwidth]{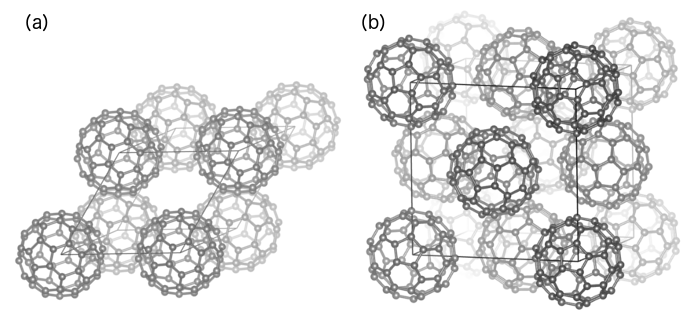}
\caption{The two types of unit cells used to model the ground state
of fullerenes and thus extract vdW-DF-cx characterizations of sublimation
energies $E_b$ of the molecular crystals. Panel (a) shows the 
fundamental organization, which is fcc type.  Panel (b) shows an example 
of a 4-C$_{60}$ super-cell structure, denoted Fm$\bar{3}$ mix for short, 
that represents an improved approximation for describing the ground-state 
of C$_{60}$ crystals.
\label{fig:C60unitcell}}
\end{figure}

Here we will show that the vdW-DF-cx functional predicts the geometries of 
fullerene solids in agreement with what is known experimentally about them. 
Other evidence for the accuracy of vdW-DF-cx is presented in Appendix B and 
in Refs.~\cite{TT,Rangel16,AAA,BBB,CCC,EHL,DDD,EEE,FFF}.

Figure \ref{fig:C60axis} shows the atom structure of C$_{60}$ and the high-symmetry 
axis (dark gray line) going out of a pair of hexagonal facets; C$_{60}$ also has a
secondary axis (light gray line) going through a pair of pentagonal facets. 
The initial coordinates of C$_{60}$, C$_{70}$, C$_{76}$, and $C_{80}$ 
fullerenes are taken from the supplemental material of Ref.~\cite{DTomanek14}.
C$_{60}$ and C$_{70}$ have $I_h$ and $D_5$ symmetry, respectively.
Since C$_{76}$, and $C_{80}$ have isomers, here we focus on C$_{76}$ and $C_{80}$ 
with $D_2$ symmetry.

\setcounter{table}{0}
\renewcommand{\thetable}{A1}
\begin{table*}
\caption{vdW-DF-cx results for primitive-cell lattice structures 
when permitting unconstrained unit-cell relaxations. They are all slightly 
distorted fcc as reflected in the lattice constants ($a, b, c$) and unit-cell 
solid angles ($\alpha, \beta, \gamma$). 
\label{tab:buckyball}
}
\begin{center}
\begin{tabular*}{0.94\textwidth}{@{\extracolsep{\fill}}llccc}
\hline \hline
Molecule & Lattice system & $a/b/c$ (\AA)  & $\alpha$/$\beta$/$\gamma$ ($^{\circ}$) 
& $V$ (\AA$^3$) \\ 
\hline
C60-Ih   & Triclinic  & 13.92/14.15/14.14  &  91.2/90.8/90.9   & 696 \\
C70-D5h  & Orthorhombic & 16.26/13.94/14.33 & 90.0/90.0/90.0  & 812 \\
C76-D2   & Orthorhombic & 16.95/15.36/13.91 & 90.0/90.0/90.0  & 905 \\
C84-D2   & Triclinic  & 16.09/15.76/15.34 & 90.3/88.4/90.0  & 972 \\
\hline
\end{tabular*}
\end{center}
\end{table*}

Panel (a) of Fig.\ \ref{fig:C60unitcell} shows the primitive (one molecule) 
and super-cell (four molecules) geometries that we have used to 
model the fullerene crystals, as illustrated with C$_{60}$.  We assume that 
fullerenes are in crystal structures that are variations of fcc. Cohesive 
energies are extracted for (super-cell) geometries that have been 
fully relaxed with the consistent-exchange vdW-DF-cx version, using the BFGS 
quasi-newton algorithm as available in variable-cell calculations ('vc-relax') 
with the \textsc{quantum espresso} package. We find no observable deformation 
of the individual fullerenes in any of the approximate-ground-state crystals 
structures that we have studied.

Table \ref{tab:buckyball} reports the details of fully unrestrained vdW-DF-cx 
charactization of the optimal structure of fullerenes, when forced into a 
primitive cell (panel (a) of Fig.\ \ref{fig:C60unitcell}). Unconstrained 
relaxation was chosen, because we do not, except for the C$_{60}$ crystal, have 
access to experimental information about alignment of fullerene symmetry axis 
and of the fullerene crystal. This vdW-DF-cx characterization yielded the 
following identification of the nature of optimal structures: triclinic 
(rhombohedral) crystal symmetry for C$_{60}$ and C$_{84}$ (for C$_{70}$ 
and C$_{76}$). From these structures, we extracted the vdW-DF-cx results
for the sublimation energies $E_b$ and for the wall-to-wall separations $d_{ww}$ 
(estimated as the distance to the nearest vertex or bond or facet on one 
molecule to the corresponding motif on the neighboring molecules). These results 
have been reported in the main text.

The C$_{60}$ crystal motivates further theoretical characterizations because 
there is experimental data on structure \cite{David92}. We note that while our 
unconstrained relaxation (modeling a primitive cell) predicts a triclinic 
structure, the actual structure deviation is small (see 
Table \ref{tab:buckyball}). This difference is, in fact, expected. The ground 
state should have two different alignments of the symmetry axis relative to the 
crystal directions \cite{David92}.  In choosing a primitive modeling, we are, 
on principle grounds, prevented from fully representing the actual C$_{60}$ 
crystal ground state.

For a deeper discussion, we consider the role of the molecular orientation 
in the C$_{60}$ crystal both in a primitive cell containing one molecule and in
an improved modeling relying on 4 molecules per unit cell; the second modeling
approach is illustrated in panel (b) of Fig.\ \ref{fig:C60unitcell}.  The 
C$_{60}$ crystal undergoes a phase transition at 260K. Above that 
temperature, all molecules can be considered equivalent, having free 
rotations, in a fcc primitive cell with one C$_{60}$ molecule. Below 
this temperature, however, the C$_{60}$ crystal is still fcc, 
but then in a super-cell configuration of unknown size. There is 
no free rotation but a forced alignment of the major symmetry 
axis with the [111] direction of the fcc crystal cell. One can 
experimentally assign a rotational-angle value, $\phi$, for 
molecules in the low-temperature systems \cite{David92}. However,
the alignment must vary over the fcc-type super cell (of unknown size)
which has a mixture of alignments: 15 \% molecules in a 
rotational configuration R$38$ and about 85 \% in R$98$. 

\begin{figure}
\centering
\includegraphics[width=0.9\columnwidth]{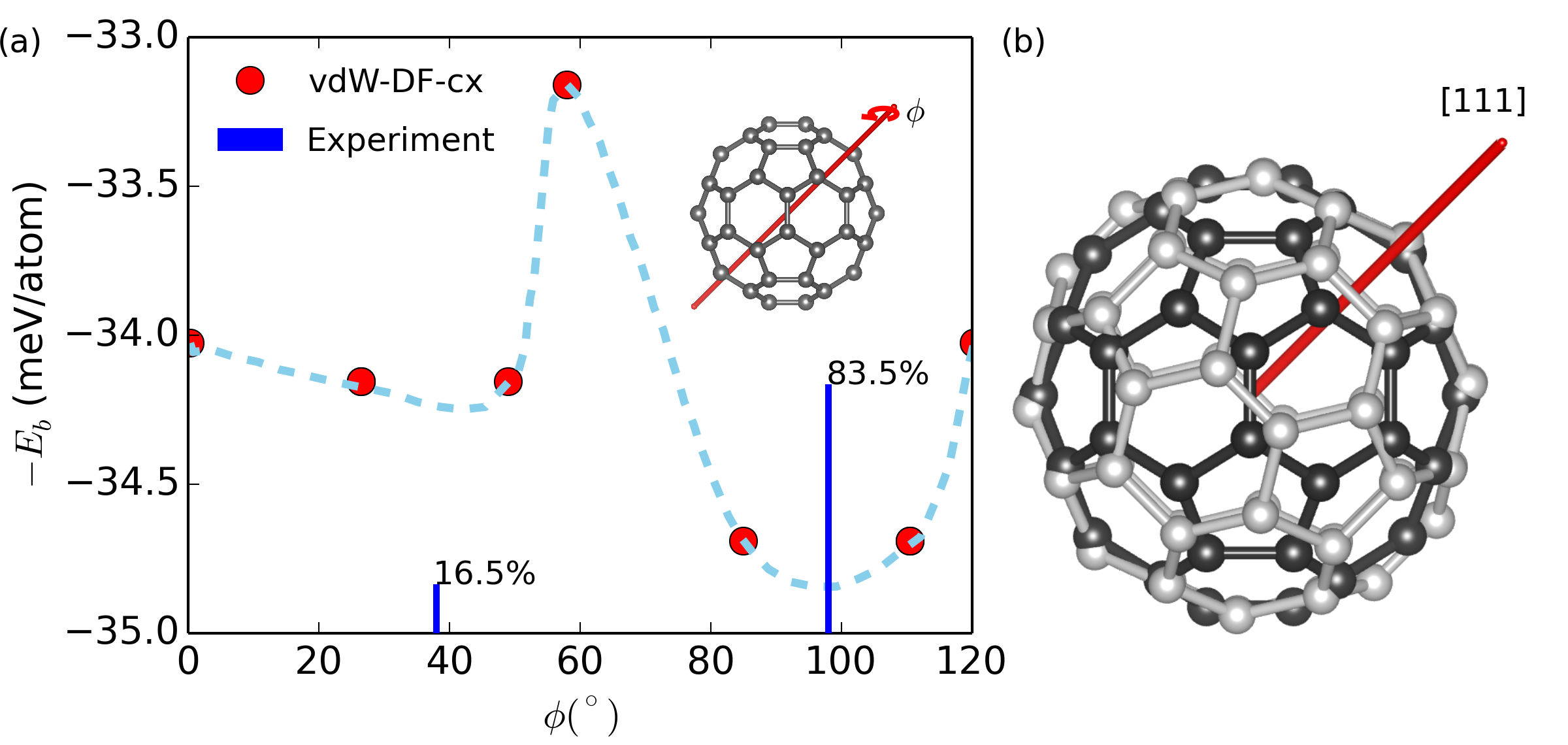}
\caption{Schematic of our test of vdW-DF-cx reliability in describing the
internal C$_{60}$ structural organization. These tests use a one-molecule cell,
but track the role of different rotations $\phi$ of the C$_{60}$ around the major 
symmetry axis.  Panel (a) shows vdW-DF-cx results for the energy variation of
such states, denoted R$\phi$;  The pair of vertical blue lines 
identifies the two experimentally observed, optimal rotational states 
that are both found in the ground state \cite{David92}.
Panel (b) contrasts the atomic structure having such alignment of the [111] 
crystal axis with the major (black atoms and bonds) against that of
alignment with the minor (gray atoms and bonds) symmetry axis. 
\label{fig:C60}}
\end{figure}

\renewcommand{\thetable}{A2}
\begin{table*}
\caption{Geometries and binding energies of meta-stable C$_{60}$ crystals as 
optimized in vdW-DF-cx with constraints. Here, for the single-molecule cell, 
we force the major C$_{60}$ symmetry axis to be aligned with the crystal axis 
and identify meta-stable configurations, denoted Fm$\bar{3}$ R$\phi^{\circ}$ 
(red dots in Fig.\ \ref{fig:C60}), with specific rotation $\phi$ values.  For 
the 4-molecule conventionnal unit cell studies, we list all meta-stable 
configurations that we have found emerging from a fcc starting point 
while permitting all four rotational angles to differ. The low-energy structure, 
`Fm$\bar{3}$ R$0^{\circ}$', has the same value for all four angles.  The 
low-energy structure `Fm$\bar{3}$ mix' is short for a four-molecule super cell
in rotational configuration R$9^{\circ}$R$23^{\circ}$R$37^{\circ}$R$116^{\circ}$.
\label{tab:C60}
}
\begin{center}
\begin{tabular}{llccccc}
\hline \hline 
 Symmetry & Lattice system & $a/b/c$ (\AA)  & $\alpha$/$\beta$/$\gamma$ ($^{\circ}$) & $V$ (\AA$^3$) & $d_{ww}$ (\AA) &
  $E_b$ (meV/atom) \\
\hline
\multicolumn{7}{c}{Experiment~\cite{David92}} \\  
 0.835 Fm$\bar{3}$ R$98^{\circ}$  & \multirow{2}{*}{Cubic}  & \multirow{2}{*}{14.04}  &  & \multirow{2}{*}{692} &  & \\
 +0.165 Fm$\bar{3}$ R$38^{\circ}$ &                         &                         &  &      &  & \\
\hline
\multicolumn{7}{c}{vdW-DF-cx; primitive cell } \\  
Fm$\bar{3}$ R$0^{\circ}$ & Cubic & 14.10/14.10/14.10 & 90.0/90.0/90.0 & 701 & 3.06 & 34.0\\
Fm$\bar{3}$ R$27^{\circ}$ & Rhombohedral & 14.08/14.08/14.08 & 90.2/90.2/90.2 & 698 & 3.11 & 34.1\\
Fm$\bar{3}$ R$49^{\circ}$ & Rhombohedral & 14.08/14.08/14.08 & 90.2/90.2/90.2 & 698 & 3.11 & 34.1\\
Fm$\bar{3}$ R$58^{\circ}$ & Rhombohedral & 14.14/14.14/14.14 & 89.9/89.9/89.9 & 707 & 3.09 & 33.2\\
Fm$\bar{3}$ R$85^{\circ}$ & Rhombohedral & 14.07/14.07/14.07 & 90.4/90.4/90.4 & 697 & 3.11 & 34.7\\
Fm$\bar{3}$ R$111^{\circ}$ & Rhombohedral & 14.07/14.07/14.07 & 90.4/90.4/90.4 & 697 & 3.11 & 34.7\\
\hline
\multicolumn{7}{c}{vdW-DF-cx; 4 molecules/unit cell} \\  
Fm$\bar{3}$ R$0^{\circ}$  & Cubic      & 13.99/13.99/13.99 & 90.0/90.0/90.0 & 684 & 2.98  &  36.6\\
Fm$\bar{3}$ mix 
& Rhombohedral & 13.96/13.96/13.96 & 90.7/90.7/90.7 & 679 & 2.96 & 36.8 \\
\hline
\end{tabular}
\end{center}
\end{table*}

Figure \ref{fig:C60} summarizes the additional structure exploration that we 
have done to test the ability of vdW-DF-cx to characterize the structural 
motifs of the C$_{60}$ ground state. The figure shows the sublimation-energy 
variation that results in a single-molecule unit-cell modeling as we vary the 
alignment of the major symmetry axis with the [111] crystal axis, all in an 
fcc structure; since the major axis has a three-fold symmetry, it is only 
necessary to explore constrained relaxations in the range $0 < \phi < 120$.
Tracking the relaxations in vdW-DF-cx, we thus identify a set of meta-stable
configurations, red dots, with specific rotations but with a range of
structural symmetries, as further described in Table \ref{tab:C60}. This class 
of metastable fcc structures differs qualitatively from the previously 
mentioned C$_{60}$ triclinic structure, which, instead, has an alignment with 
the secondary symmetry axis in C$_{60}$. However, as we show later, the energy 
differences are very small.

\renewcommand{\thetable}{A3}
\begin{table*}
\caption{Comparison of geometries and binding energies of C$_{60}$ crystals in 
metastable cubic/rhombohedral structures (in which the [111] crystal axis is 
kept aligned with major symmetry axis) and in alternative triclinic structures  
(in which [111] is found to be aligned with the secondary C$_{60}$ symmetry 
axis). The former is a characteristics of the experimentally observed ground 
state, while the latter is what emerges in unconstrained relaxations in the 
one-molecule primitive cell.
}
\label{tab:C60var}
\begin{center}
\begin{tabular}{llccccc}
\hline \hline 
 Symmetry & Lattice system & $a/b/c$ (\AA)  & $\alpha$/$\beta$/$\gamma$ ($^{\circ}$) & $V$ (\AA$^3$) & $d_{ww}$ (\AA) &
  $E_b$ (meV/atom) \\
\hline
\multicolumn{7}{c}{Experiment~\cite{David92}} \\  
 0.835 Fm$\bar{3}$ R$98^{\circ}$  & \multirow{2}{*}{Cubic}  & \multirow{2}{*}{14.04}  &  & \multirow{2}{*}{692} &  & \\
 +0.165 Fm$\bar{3}$ R$38^{\circ}$ &                         &                         &  &      &  & \\
\hline
\multicolumn{7}{c}{vdW-DF-cx; primitive cell } \\  
 -        & Triclinic  & 13.92/14.15/14.14  &  91.2/90.8/90.9   & 696 & 3.05  &  34.8\\    
Fm$\bar{3}$ R$0^{\circ}$ & Cubic & 14.10/14.10/14.10 & 90.0/90.0/90.0 & 701 & 3.06 & 34.0\\
Fm$\bar{3}$ R$111^{\circ}$ & Rhombohedral & 14.07/14.07/14.07 & 90.4/90.4/90.4 & 697 & 3.11 & 34.7\\
\hline
\multicolumn{7}{c}{vdW-DF-cx; 4 molecules/unit cell} \\  
 -                        & Triclinic  & 13.90/14.08/14.04 & 91.2/91.0/90.8 & 687 & 2.93 & 36.1 \\
Fm$\bar{3}$ mix 
& Rhombohedral & 13.96/13.96/13.96 & 90.7/90.7/90.7 & 679 & 2.96 & 36.8 \\
\hline
\end{tabular}
\end{center}
\end{table*}

The dashed line in the left panel of Fig.\ (\ref{fig:C60}) represents a guide to 
the eye among those meta-stable configurations. We assume that a full exploration 
would stabilize major-axis configurations at more rotational values, when pursued 
at a super-cell size that reflects the actual ground state. If we furthermore 
take the dashed line as an approximation for how such additional local minima 
would be distributed in energies, then we can expect optimal rotational values 
at around $\phi\approx 40^{\circ}$ and $\phi\approx 100^{\circ}$. It is 
heartening that these angles coincide with those that emerge as most prevalent in
the mixture description obtained in the experimental characterization of 
C$_{60}$, evident as vertical blue lines in Fig.\ \ref{fig:C60}. 

Table \ref{tab:C60} also reveals that structural optimization in the super cell
indicates a very small preference for mixing different molecular rotations.
Here again the relaxation is constrained to the experimentally observed 
major-axis alignment. In a super cell, however, we can allow 
individual molecules to relax to different orientation values. The structure
identified as `Fm$\bar{3}$ mix' is an example of an energetically favorable
such meta-stable configuration. Like the actual but unknown ground-state super 
cell,\cite{David92} this structure is characterized by having a mixture of 
molecular rotations. 

Finally, Table \ref{tab:C60var} lists the sublimation energies that 
arises when the [111] crystal axis (red line) is assumed to align with either 
the major or the secondary symmetry axis for the C$_{60}$ crystal. For a 
single-molecule modeling, and among the cases with major-axis alignment, we 
find a best case, Fm$\bar{3}$ R$111^{\circ}$, with a sublimation energy that 
lies with 0.1 meV/atom of that of the triclinic structure (with the 
secondary-axis alignment). Also, although the energy differences are still 
very small, the $E_b$ ordering is reversed when instead we consider the best 
four-molecule super celll case (with correct alignment), denoted 
`Fm$\bar{3}$ mix'.

Overall, we find that the vdW-DF-cx is able to reflect the known structural 
motifs of the C$_{60}$ crystal (although the C$-{60}$ ground state is not 
fully known): (a) preference for a fcc-type super-cell configuration with a 
mixture of rotational angles, (b) a preference for having predominently a 
$\phi=100^{\circ}$ rotational state, and (c) a per-molecule volume value, 
which for the most-favorable super-cell representation, lies within 2 percent 
of the experimentally observed value, 692 {\AA}$^3$. We take this vdW-DF-cx 
progress as an indication that it can be used to predict the binding 
structures of the set of investigated nano-structured materials and that it 
is a good starting point for exploring energy scaling laws. 

\section{Asymptotic binding in vdW-DF-cx}

\renewcommand{\thetable}{B1}
\begin{table*}
\caption{Asymptotic van der Waals interaction coefficients $C_6$
and molecular-crystal sublimation (or cohesion)
energies $E_b$ for fullerenes. The van der Waals
interaction coefficients are in atomic units
(hartree for energy, bohr  for distances) while $E_b$
is reported in meV per carbon atom. The vdW-DF-cx results
for $E_b$ are listed for the primitive
(one fullerene per) cell studies (see Appendix A).
The experimental results~\cite{MMoalem95,BBrunetti97} are enthalpies of 
sublimation without thermal corrections.
}
\label{tab:C6}
\centering
\begin{tabular}{l|cc|cc|c|c}
\hline \hline
    & \multicolumn{2}{c|}{vdW-DF-cx}        & \multicolumn{2}{c|}{TDHF~\cite{JCP13}} & vdW-DF-cx & Experiment \\
    & $C_6/10^3$ & $C_6/{\cal N}^2$ & $C_6/10^3$ & $C_6/{\cal N}^2$ & $E_b/{\cal N}$    &  $E_b/{\cal N}$   \\
\hline
C$_{60}$ & 55.2 & 15.33 & 100.1 & 27.80 & 35.0 &  25$\sim$32 \\
C$_{70}$ & 74.8 & 15.26 & 141.6 & 28.90 & 32.9 &  24$\sim$30\\
C$_{76}$ & 88.1  & 15.25 & -     & - & 32.2 &  26$\sim$27 \\
C$_{84}$ & 107.3  & 15.20 & 207.7 & 29.44 & 31.6 & 27$\sim$34 \\
\hline
\end{tabular}
\end{table*}

Here we will show that vdW-DF-cx predicts equilibrium binding energies of 
fullerene solids in agreement with experiment, even though, as anticipated in 
Ref.~\cite{PRL12} it is not accurate for the asymptotic interaction of 
a fullerene pair.

We have extracted the C$_6$ values that correspond to an asymptotic
vdW-DF-cx description, following prior discussions of the nature of the 
vdW-DF binding \cite{Dion04,kleis08}. For the finite fullerene structures, this 
leads to a determination of the C$_6$ coefficients using Eq. (5)-(7) of 
Ref.~\cite{per11}. The following information is presented to permit a discussion 
of differences in the asymptotic description from the hollow-shell 
model \cite{JCP13}.

In the vdW-DF method, we work with a local-field susceptibility $\alpha$ and 
corresponding external-field susceptibility 
$\alpha_{\rm ext}=\alpha/(1+4\pi\alpha)$. This susceptibility 
(or plasmon propagator) depends on two spatial coordinates but can be 
represented in a gradient expansion. When investigating the asymptotic 
interactions, the relevant limit is \cite{Dion04,kleis08}:
\begin{equation}
\alpha_{\rm ext, assym}^{\rm{vdW-DF}}(iu,r) \to  \frac{n(r)}{u^2+[9q_0(r)^2/(8\pi)]^2} \, .
\label{eq:suscept}
\end{equation}
This susceptibility limit is directly set by the inverse length scale 
$q_0$ that enters in the specification of the local plasmon 
dispersion \cite{Dion04,kleis08,PerHyldgaardPRB14}.  We compute this 
susceptibility limit from the electron-density variation $n(\mathbf{r})$ 
that we have established in the underlying (full) vdW-DF-cx calculations.  

From the approximation Eq.\ (\ref{eq:suscept}) we determine, in turn, the 
asymptotic vdW-DF-cx description of nanostructure interaction from a 
Casimir-Polder expression of the molecular C$_6$ coefficients, using a 
numerical imaginary-frequency integration. The result is a description 
similar to Eq. (17) of Ref.~\cite{Dion04}. We note that this $C_6$ limit 
is not an exhaustive representation of the full, regular vdW-DF-cx 
calculations, for reasons explained in Refs.\ \onlinecite{Dion04,kleis08,
per11,PerHyldgaardPRB14}.

Table \ref{tab:C6} reports our numerical extraction of such vdW-DF-cx 
based C$_6$ coefficients for fullerenes with various number of carbon 
atoms ${\cal N}$, here contrasted with TDHF-based values cited in Table 3 
of the main text.  In general, the nonlocal-correlation part of 
vdW-DF \cite{Dion04} (same as in vdW-DF-cx) leads to good C$_6$ values for 
small-to-medium sized molecules \cite{VV101}. However, for the hollow 
fullerenes, the vdW-DF-cx values are about half the size of the results of 
the shell-model analysis based on TDHF calculations \cite{JCP13}. Also, 
the vdW-DF-cx results for C$_6$/${\cal N}^2$ is nearly independent of 
${\cal N}$. Thus the vdW-DF-cx descriptions does not reflect the C$_6$ 
nonadditivity that is expressed in the TDHF-based C$_6$ 
description \cite{JCP13} and hollow-sphere model.  

The vdW-DF-cx functional slightly overestimates the nanostructure binding 
energies $E_b$, but it is still useful for mapping the energy scaling as it 
is accurate on structure characterizations (Appendix A and 
Refs.~\cite{KBerland14-1,Rangel16,BBB,DDD}), as well as for nanostructure 
energy differences~\cite{KBerland14-1,TT,CCC,DDD}. Table B1 also reports a
comparison of the vdW-DF-cx results for fullerene sublimation energies $E_b$
and raw experimental observations (no thermal correction); 
Table \ref{table1} reports the comparison when the finite-temperature effect 
is removed from the experimental values.

For the C-PAH dimers in the sandwich or 
AA configuration, CCSD(T)-based best estimates of the binding 
energy per atom for benzene, naphthalene, and anthracene~\cite{Silva16} 
are 13.2, 17.8, and 21.1 meV per carbon atom, versus the vdW-DF-cx 
values of 17.4, 22.4, and 25.0 meV per carbon atom, respectively. 
For the corresponding benzene, naphthalene, and anthracene 
molecular crystals, the measured supplimation energies are 
22, 41, and 40 meV per carbon atom, versus 
the vdW-DF-cx values of 25, 46, and 44 meV per carbon atom~\cite{TT}. 
Thus vdW-DF-cx captures the right trends with increasing particle 
size (being in these cases about 4 meV per carbon atom 
higher than the best available estimates). 

Moreover, vdW-DF-cx is accurate in first-principle
characterizations of nanostructure-energy differences. 
It has proven useful for the description of both 
elastic deformations energies \cite{KBerland14-1,AAA,EHL,EEE} 
and lattice vibrations \cite{EHL,CCC,DDD,EEE}. For example,
it can accurately reproduce the measurement of all 
all libration modes in the naphthalene molecular 
crystal, characterizing the phonon dispersion 
to within 1 meV (without any experimental input on
the molecular-crystal structure) \cite{EEE}. 

Some of us have previously (independently) discussed that it is 
possible for a density functional to fail in the asymptotic description 
but still be accurate at binding separations \cite{PerHyldgaardPRB14,
PengPRX}. The vdW-DF-cx description of the fullerene crystals gives in 
example: Appendix A shows that vdW-DF-cx does give an accurate description 
of structural motifs in fullerene crystals at binding separation even if 
vdW-DF-cx is not accurate for (and does not give nonadditive) C$_6$ 
coefficients, Table \ref{tab:C6}. We also note that the vdW-DF-cx is 
nonadditive in a different sense, namely in its description of the 
nonlocal correlation interaction at binding separations \cite{Mapping17}.

\end{document}